# Hidden aspects of the Structural Theory of chemistry: The MC-QTAIM analysis reveals "alchemical" transformation from a triatomic to a diatomic structure

Mohammad Goli and Shant Shahbazian[*]

*Faculty of Chemistry, Shahid Beheshti University, G. C. , Evin, Tehran, Iran, 19839, P.O. Box 19395-4716.*

Tel/Fax: 98-21-22431661

E-mail:
(Shant Shahbazian) chemist_shant@yahoo.com

[*] Corresponding author

# Abstract

The Structural theory of chemistry introduces chemical/molecular structure as a combination of relative arrangement and bonding patterns of atoms in molecule. Nowadays, the structure of atoms in molecules is derived from the topological analysis of the quantum theory of atoms in molecules (QTAIM). In this context a molecular structure is varied by large geometrical variations and concomitant reorganization of electronic structure that are usually taking place in chemical reactions or under extreme hydrostatic pressure. In this report a new mode of structural variation is introduced within the context of the newly proposed multi-component QTAIM (MC-QTAIM) that originates from mass variation of nuclei. Accordingly, *XCN* and *CNX* series of species are introduced where *X* stands for a quantum particle with a unit of positive charge and a variable mass that is varied in discrete steps between the masses of proton and positron. Ab initio non-Born-Oppenheimer (non-BO) calculations are done on both series of species and the resulting non-BO wavefunctions are used for the MC-QTAIM analysis revealing a triatomic structure for the proton mass and a diatomic structure for the positron mass. In both series of species a critical mass between that of proton and positron mass is discovered where the transition from triatomic to diatomic structure takes place. This abrupt structural transformation has a topological nature resembling the usual phase transitions in thermodynamics. The discovered mass induced structural transformation is a hidden aspect of the Structural theory which is revealed only beyond the BO paradigm when nuclei are treated as quantum waves instead of clamped point charges.

# Keywords

# 1. Introduction

The original notion of the molecular/chemical structure, as the relative arrangement and bonding pattern of atoms of a molecule, has been probably the most fundamental concept in chemistry since the advent of the Structural theory by Kekule, Couper and Butlerov over one and half century ago.[1-3] Since the introduction and subsequent development of the Lewis electronic theory of atoms and molecules almost a century ago,[4] a new dimension namely, electrons distribution, has also been incorporated into the very notion of the molecular structure. The notion of molecular structure did not cease to evolve, after all subsequent experimental and theoretical advances, e.g. the emergence of X-ray crystallography and quantum chemistry. Nowadays, it is generally believed that once the equilibrium nuclear configuration and the electronic structure of a molecule are known, essentials to derive molecular structure is at hand.[5] Currently, apart from sophisticated experimental methods,[6,7] computational quantum chemistry is also a reliable source for deriving equilibrium nuclear configurations.[8-10] Besides, the resulting ab initio electronic wavefunctions are used as "input" to those quantum chemical methodologies, e.g. natural bond orbital method and its ramifications,[11-14] that aim to obtain a "chemical"/real space picture of electronic structure. Alternatively, the quantum theory of atoms in molecules (QTAIM) that directly deciphers the "atoms in molecules" (AIM) structure from the electronic wavefunctions is a useful methodology which in its approach is close to the notion of the original Structural theory.[15-17] Through the combination of elements of these two levels of *chemical organization*, i.e. electronic and the AIM structures, one hopes that a detailed and consistent picture of the molecular structure emerges.

However, all these quantum chemical methodologies and the emerging image of molecular structure have roots in the familiar Born-Oppenheimer (BO) paradigm.[18] In this paradigm electrons are considered as quantum waves whereas nuclei are treated as clamped particles acting as the source of the Coulombic electric field that bounds electrons. The "dual" role of electrons and nuclei is attributed to their large mass difference that justifies an adiabatic viewpoint. Recent advances in non-BO ab initio computational methods bypass this paradigm demonstrating that nuclei may also be treated as quantum waves from outset of calculations without any need to adiabatic picture even at an intermediate stage.[19-28] Then, the question naturally emerges that how the familiar notion of molecular structure is derivable from a non-BO wavefunction. To answer this question, novel quantum chemical methodologies must be developed being capable of using non-BO wavefunctions, instead of adiabatic electronic wavefunctions, as input to extract essentials of molecular structure. In contrast to some primary progress in this direction,[29-41] currently, such novel non-BO methodologies are in their infancy and it is not clear whether they will survive in long term as reliable sources to derive essentials of molecular structure.[18,42-46] A more modest strategy is extending the known BO-based methodologies to the non-BO realm; the use of various "extended population analysis" methodologies using non-BO wavefunctions as input is a prime example.[47-63]

The recently proposed multi-component quantum theory of atoms in molecules (MC-QTAIM) also serves as an example of methodologies that aim to bypass the BO paradigm, unraveling the AIM structure from non-BO wavefunctions.[64-70] While the orthodox QTAIM is confined within the BO paradigm,[15-17] the MC-QTAIM is capable of revealing the AIM structure from both the BO and non-BO wavefunctions unifying the AIM analysis of both realms. In fact, it has been demonstrated that the orthodox QTAIM is just the "asymptote" of the MC-QTAIM when the mass of nuclei tends toward infinity.[66] On the other hand, the MC-QTAIM is also capable of revealing the AIM structure of exotic species containing new fundamental particles beyond the electrons and the familiar nuclei that the positronic and muonic molecular systems are primary examples.[71-74] Thus, the unification domain of the MC-QTAIM methodology extends further promising the *extension* of the Structural theory beyond its classic territory.[69] Particularly, this is a promising and fresh field of research since high quality ab initio wavefunctions are now accessible for many exotic species.[75-81] On the other hand, one of the main traits of the MC-QTAIM analysis is its mass dependence namely, the masses of the constituent bodies of the molecular system are directly involved in the underlying formalism. Accordingly, both the Gamma density that is used for the topological analysis and unraveling the boundaries of atomic basins, and the property densities yielding properties of atomic basins, are explicitly mass dependent (*vide infra*).[66] One of manifestations of this mass-dependency is the observation of distinct atomic basins for each of the hydrogen isotopes.[64,74]

Another opportunity, which is scrutinized in this study, is *direct* observation of the variation of molecular structure upon the mass variation of constituent bodies. In order to unravel this variation, two series of systems including $XCN$ and $CNX$ series are considered where in both series $X$ stands for a quantum particle with a unit of positive charge and a variable mass that is varied in discrete steps (*vide infra*) between the masses of proton and positron, $m_{proton} \approx 1836 m_e$ and $m_{positron} = m_e$, respectively. Both extremes correspond to well known molecular structures. In the case of proton, both $HCN$ and $CNH$ are linear structures containing three atomic basins each encompassing one of the nuclei.[15] On the other hand, in the case of positron one is faced with positronic cyanide, $CN^-, e^+$ which is a diatomic positronic species with two atomic basins corresponding to the two nuclei whereas the positron, unable to shape its own atomic basin, is distributed unevenly within these two baisns.[72] Intuitively, one expects that with the mass variation of the positively charged particle (PCP) from one extreme to the other, a structural transition must be somehow taken place from a triatomic to a diatomic structure. As is demonstrated in the rest of this contribution, this "mass dependent" structural transition has a topologically *abrupt* nature within context of the MC-QTAIM and there is a *critical mass* of the PCP witnessing this sudden/catastrophic transformation.

## 2. A brief survey on the computational procedures and the mass dependence of the Gamma density

The non-BO ab initio computational procedure used in this study is the multi-component Hartree-Fock (HF) method developed within the context of the Nuclear-Electronic-Orbital theory termed NEO-HF,[23] as implemented in the NEO computer code that is now part of the GAMESS suite of programs.[82] In the NEO-HF method, which is based on the multi-component Hartree-Fock equations, the non-BO wavefunction is approximated as a product of Slater determinants (assuming constituent particles as fermions).[23] In the present study in both series of species the wavefunction is the product of a Slater determinant, constructed from electronic orbitals, and an orbital describing the PCP all determined from the SCF procedure employing the mean potential field of the multi-component Hartree-Fock equations. As detailed elsewhere recently,[70] new capabilities have been added to the original NEO code including the mass variation of the PCP as well as using a "hybrid" basis set consists of the usual nuclear centered and "mobile" functions. The concept of mobile function is particularly vital in considering cases that Gaussian basis sets with pre-fixed exponents and positions are useless. This is mainly the case in this study considering PCPs with non-standard masses so they are scrutinized in this section in details.

In mobile part of the basis set all variables of Gaussian functions which includes the exponents, positions and linear coefficients were optimized variationally during the SCF procedure of the NEO-HF method. Throughout the ab initio calculations carbon and nitrogen nuclei were treated as clamped particles whereas electrons and the PCP were treated as quantum waves containing kinetic energy operators in the multi-component HF equations. In present study all the centers of the basis functions were placed on an axis going through both the clamped nuclei, and their position variations were restricted only on this axis during the optimization procedure. To describe the electrons around the clamped nuclei, the standard nuclear centered cc-pVDZ electronic basis set was placed on the clamped nuclei.[83,84] On the other hand, the PCP's orbital in the $XCN$ series has been composed of three mobile s-type Gaussian functions each with a different center, one placed between the two clamped nuclei and the two others each behind each of the clamped nuclei. This arrangement of Gaussian functions was used as an "initial guess" in the initial step of the NEO-HF calculations and then varied during the optimization process. For the $CNX$ series just a single s-type Gaussian function was placed behind the clamped nitrogen nucleus and used in the initial step of the NEO-HF calculations. This strategy was employed since the computational experiences demonstrated that in most cases the variationally optimized SCF coefficients of the three s-type Gaussian functions describing the PCP's orbital prefer the $XCN$ configuration; only the SCF coefficient of the Gaussian function behind the carbon clamped nuclei is non-zero except the positronic cyanide, $CN^-, e^+$. In order to describe electronic orbitals surrounding the above mentioned PCP associated mobile functions in both series, nine s-type Gaussians functions were grouped into three subsets as electronic mobile functions, each subset containing three Gaussians, and were placed at three different centers. In the $XCN$ series these three centers were fixed at the center of the three mobile functions describing the PCP's orbital. For the $CNX$ series one subset was fixed at the center of the single s-type Gaussian function describing the PCP's orbital while the two others were initially placed one between the clamped nuclei and one behind the clamped carbon nucleus and treated as "ghost" centers (sometimes also called banquet/Bq nuclei). Though these two ghost centers are not in practice describing electronic distribution around the PCP, they are used to make the description of electronic structure in the $CNX$ series as flexible as that of the $XCN$ series. At the next stage during the NEO-HF calculations of each species in both series in addition to the usual linear optimization of the SCF coefficients of the hybrid basis set,[23] the inter-nuclear distance of the clamped nuclei and the centers and exponents of the mobile functions were all optimized with a non-linear non-gradient optimizer added externally to the NEO code. At the final stage of ab initio calculations the gradient of energy was computed explicitly to ensure the precision of the external optimizer; the root mean square of forces operative on the nuclei and the PCP was always less than $10^{-5}$ Hartree/Bohr in all considered species while the maximum force was always less than $10^{-4}$ Hartree/Bohr. The designed procedure is a combination of the orthodox NEO-HF, which is usually used with fixed-center basis sets with pre-defined exponents,[23] and the fully variational multi-component Hartree-Fock method that optimizes all variables of a basis set.[26,27,64-68,70] In the series of the $XCN$ species the mass of the PCP was varied in discrete steps including: $m_X = m_e$, $25m_e$, $50m_e$, $60m_e$, $65m_e$, $70m_e$, $75m_e$, $80m_e$, $85m_e$, $90m_e$, $95m_e$, $100m_e$, $200m_e$, $600m_e$, $1000m_e$, $1400m_e$, $1836m_e$, while in the $CNX$ series these include: $m_X = 25m_e$, $50m_e$, $100m_e$, $200m_e$, $300m_e$, $350m_e$, $385m_e$, $400m_e$, $405m_e$, $410m_e$, $415m_e$, $425m_e$, $500m_e$, $600m_e$, $1000m_e$, $1400m_e$, $1836m_e$. Apart from the PCPs with $m_X = m_e$, $200m_e$, $1836m_e$ that match to the masses of the positron, and almost to the muon and the proton, respectively, the other members of the "mass spectrum" are fictitious particles just employed to reveal a semi-continuous view of patterns emerging in the MC-QTAIM analysis from the mass variation.

The details of the MC-QTAIM formalism and associated computational analysis have been fully disclosed elsewhere and are not reiterated here.[64-70] However, because the mass dependent topological transitions are of main focus in this study, the mass dependence of the Gamma density is reviewed briefly. Since there are two types of quantum particles in the considered systems, i.e. electrons and the PCP, the proper Gamma field is: $\Gamma^{(2)}(\vec{q}) = \rho_e(\vec{q}) + (m_e/m_X)\rho_X(\vec{q})$ where $\rho_e(\vec{q}) = 14\int d\tau'_e \Psi^*\Psi$ is the one-particle density of electrons while $\rho_X(\vec{q}) = \int d\tau'_X \Psi^*\Psi$ is the one-particle density of the PCP.[67,70] In these equations $d\tau'_t$, $t = e, X$ implies summing over spin variables of all quantum particles and integrating over spatial coordinates of all quantum particles except one arbitrary particle belonging to the subset of electrons, denoted by subscript $e$, or the PCP, denoted by subscript $X$. Since $\Psi$ is a non-BO wavefunction it is important to distinguish the one-particle density of electrons used to construct the Gamma density and the usual one-electron density introduced within the context of BO paradigm and employed in the topological analysis of the orthodox QTAIM.[15-17] The mass dependence of the Gamma density is of two types; the "explicit" mass dependence originating from the equation of the Gamma and an "implicit" mass dependence which stems from the fact that both of the one-particle densities are also mass dependent, themselves. By decreasing the mass of the PCP, in the case of explicit mass dependence, the contribution of the PCP's one-particle density increases in the combined Gamma density. However, the implicit mass-dependence dictates that a PCP with a smaller mass, which yields a less localized PCP one-particle density, is less capable of accumulating electrons. This is best exemplified in the above-mentioned extremes in both *XCN* and *CNX* series. While the proton is capable of accumulating electrons sufficiently to yield a (3, -3) CP in the gradient vector field of the Gamma density and shaping a topological atom,[85] the one-particle density of proton has a very little direct contribution to the Gamma density. On the other hand, because positron and electron have the same mass, positron's one-particle density is formally contributing equally to the Gamma density like the electrons' one-particle density but its extreme diffuseness,[72] makes it quite ineffective to accumulate electrons and shaping a topological atom. Among these two extremes there is a competition between the two opposite factors in shaping the topography of the Gamma density. Thus, a prior prediction of the *critical mass* of the PCP where the atomic basin associated to the PCP appears/disappears, without a detailed computational study, is not straightforward. The atomic properties of each atomic basin, $\tilde{M}(\Omega)$, are determined by basin integrations of the combined property density. The property densities are composed of two contributions originating from electrons, $M_{-}(\vec{q})$, and the PCP, $M_{X}(\vec{q})$, thus:

$$\tilde{M}(\Omega) = \int_{\Omega} d\vec{q} \ \tilde{M}(\vec{q}) = \int_{\Omega} d\vec{q} \ \{M_{-}(\vec{q}) + M_{X}(\vec{q})\} = M_{-}(\Omega) + M_{X}(\Omega).$$

The numerical basin integrations were done using an algorithm disclosed elsewhere and is not reconsidered.[64,72] The computed virial ratios (see Table 1), $\langle V \rangle / \langle T \rangle$, deviate from the exact value, $-2$, thus, an *ad hoc* virial scaling were done when computing atomic energies.[70,72] This stems from the "partial" optimization of the basis function; although all variables of mobile functions were fully optimized, in the case of nuclear centered function only the SCF coefficients, but not centers and exponents, were optimized. In order to guarantee the accuracy of the basin integration procedure, the numerical parameters of the basin integration were varied until the net flux integral of each basin, $\tilde{L}(\Omega) = (-1/4) \int_{\Omega} d\vec{q} \ \nabla^2 \Gamma^{(2)}(\vec{q})$, was smaller than $10^{-4}$ in atomic units. Species from both series whose basin integrations, because of numerical instabilities, do not conform to this criterion, namely $\tilde{L}(\Omega) > 10^{-4}$, were excluded from the final MC-QTAIM analysis. The accuracy of the basin integrations was double checked comparing the sum of basin properties with associated molecular properties which were computed independently from the ab initio calculations, $M_{molecule} = \sum_{\Omega} \tilde{M}(\Omega)$.

## 3. Results and Discussion

Table 1 offers some results of the NEO-HF calculations and considering the mass variation of the PCP, regular patterns are observable in the geometrical parameters and total energies of the considered species in both series. In discussing the patterns throughout the paper it is always assumed that all trends are described from $HCN/CNH$ to $CN^-, e^+$ thus the phrase "because of the decrease of the mass" is eliminated from corresponding statements. In both series the *C-N* inter-nuclear distances and particularly the mean *C-X* and *N-X* distances are ascending indicating "spatial expansion" of species containing lighter PCPs in both series. The absolute amount of the total energy is descending in both series that is in line with the observed spatial expansion. This well-known trend is rationalized considering the fact that a heavier PCP accumulates electrons in a smaller region, i.e. tighter orbits, increasing the electronic kinetic energy that dominates the total kinetic energy, $\langle T \rangle$, and also the total energy, $E_t$, through the virial theorem, $E_t = -\langle T \rangle$.[65,68,70] On the other hand, comparing congener species from both series with the same PCP mass demonstrates that the $XCN$ species are always more stable than $CNX$ species. This observation is in line with the observed "variational collapse" of the three center mobile basis set to the $XCN$ configuration in most of the mass spectrum (*vide supra*).

The MC-QTAIM analysis starts with the topological analysis of the Gamma density which its relief maps are depicted for selected species from both series in Figures 1 and 2. The topological analysis of the species in both series clearly demonstrates that one is faced with two distinct types of Molecular Graphs (MGs) when scanning the mass spectrum. In the case of the $XCN$ series the derived MGs in the mass region $85m_e \leq m_X \leq 1836m_e$ are topologically equivalent to the usual MG that is derived for $HCN$ within context of the orthodox QTAIM.[15] Three (3, -3) critical points (CPs) reveal three basins of attraction within the vector gradient field of the Gamma density which are atomic basins separated with the zero-flux surfaces going through the two (3, -1) CPs. Accordingly, the PCP, though not clamped, is capable of forming an atomic basin in this mass region. On the other hand, in the mass region $m_e \leq m_X \leq 80m_e$ the (3, -3) CP associated to the PCP disappears and corresponding species are now composed of two atomic basins. Evidently, a topological transformation takes place within the mass region $80m_e \leq m_X \leq 85m_e$ witnessing *an abrupt structural transition from a triatomic to a diatomic structure*. A similar pattern is also observed in the $CNX$ series though in this case the structural transition occurs in the mass region $405m_e \leq m_X \leq 410m_e$. Accordingly, the *critical mass* of the PCP for the structural transition is quite different in the two series revealing the interesting fact that the positive muon, $m_\mu \approx 206.8m_e$, is capable of forming its own basin in the $XCN$ series but not in the $CNX$ series. Tables 2 as well as Tables S1 and S2 in supporting information gather some quantitative results of the topological analysis including the topological indices at the (3, -3) and (3, -1) CPs. It is evident from Tables S1 and S2 that in both series of species the topological indices of (3, -3) CPs at the clamped nuclei and the (3, -1) CP in between are almost constant and relatively insensitive to the mass variation of the PCP. As is also stressed recently,[70] this is a manifestation of the principle of the *nearsightedness of the electronic matter*,[86,87] which roughly states that a perturbation induced by a variation on a specific site of a molecule damps beyond the region of perturbation when considering the electronic density as well as the property densities. Manifestly, the mass variation of the PCP, which acts similar to a perturbation, is best manifested on the topological indices computed at the PCP associated (3, -3) CP as well as the (3, -1) CP at the boundary of the *X* basin and its neighboring basin. Inspection of Table 2 reveals that in both series the amount of the Gamma density as well as the absolute amount of its Laplacian are both descending at the (3, -3) CP demonstrating that the lighter PCPs are less capable of concentrating/accumulating electrons around themselves; in order to gain a more detailed picture, Tables S3 and S4 in supporting information offer the separate contributions of electrons and the PCP to the topological indices. The length of the gradient path connecting the (3, -3) CP in the *X* basin and the (3, -1) CP at the boundary of the *X* basin and its neighboring basin is also descending in both series and tends to zero near the critical masses. On the other hand, the topological floppiness index, $TF = \Gamma^{(2)}(\vec{q}_{(3,-1)})/\Gamma^{(2)}(\vec{q}_{(3,-3)})$,[68] is ascending and approaching its limiting value, $TF = 1$, near the critical masses in both series. All these conform to the fact that the PCP associated atomic basin "shrinks" and suddenly disappears at the critical mass.

Some atomic properties derived from the basin integrations are gathered in Tables 3, 4 and 5. Inspection of Table 3 demonstrates that in both series of species the absolute amount of basin energies, electronic populations, the population of the PCP and the atomic volumes are all revealing a rapidly descending pattern for the *X* basin. These observations conform to the results of the topological analysis and the annihilation of the *X* basin below the critical mass. Interestingly, comparison of congener species from both series, those containing PCPs having the same mass, reveals that in the *CNX* series the *X* basin is always smaller, containing fewer electrons. This is rationalized taking into account the fact that nitrogen basin, which is the neighbor of the *X* basin in the *CNX* series, has a larger capacity of electron withdrawing/electronegativity than the carbon basin which is the neighbor of the *X* basin in the *XCN* series. Accordingly, based on Tables 4 and 5, in both series the atomic charges of the nitrogen basins, results from subtracting nitrogen's atomic number and its electronic population,[68] are always negative in line with its place in electronegativity scale whereas those of the carbon basins are always positive. A more detailed inspection of these tables also reveals that in the set of the nitrogen and carbon basins in both series a larger electron population always accompanied with more negative basin energy. Such trend, which recently has also been observed in the case of hydrogen basins,[70,74] demonstrates that basin energies are sensitive probes of electron transfer processes though a detailed theoretical understanding is yet missing. Also, in the *CNX* series and some of lighter species in the *XCN* series the PCP is not completely contained within the *X* basin and "leaks" into the neighboring basin. Evidently, in competition with neighboring basin the capacity of the *X* basins, containing lighter PCPs, to maintain the electrons and even the PCP itself weakens. The PCPs leakage is also observable in Tables S3 and S4 considering even the mass-scaled PCP's contribution to the Gamma density at the (3, -1) CP at the boundary of the *X* basin and its neighbor. Inspection of Tables 4 and 5 demonstrates that in both series, in line with the principle of nearsightedness, the basin that is neighbor of the *X* basin is more influenced by the mass variations of the PCP. Whereas, the pattern of property variations of the non-neighboring basins is gradual and virtually "continuous", at border of the critical mass, the neighboring basin experiences "discontinuous" property variations. These discontinuous property variations originate from the fact that below the critical mass the basin neighboring of the *X* basin completely "absorbs" the *X* basin, i.e. both its electrons and the PCP, into itself yielding a single basin; the whole process resembles the well-known discontinuous phase transitions in thermodynamics. *In this analogy the two topological structures above and below the critical mass act like two distinct phases while the* *mass of the PCP acts like the control parameter, e.g. temperature.* It is timely to emphasize that this analogy was recognized and considered in the original literature of the orthodox QTAIM.[15] However, since the formalism of the orthodox QTAIM is confined to the BO paradigm, the control parameters are only nuclear coordinates and the topological changes are solely induced by molecular geometry variations.

It is also interesting to comment on the special position of the positronic molecule among the other members of these two series since it is the only species that nitrogen basin shares the population of the PCP, Figure 3, and its atomic properties are distinctly different from those of species with $m_X = 25m_e$, Table 4. Although one does not expect a new topological structural transformation to take place between $m_X = m_e$ and $m_X = 25m_e$, evidently, the PCP's migration into the nitrogen basin starts in this mass region that seems to be the final stage of property variations in the mass spectrum. Attempts to survey this mass region were plagued by the fact that multiple "local" minima with similar energies were emerged in the "space" of variables of the mobile functions for each species. Thus, it is hard to unambiguously locate the "global" minimum for each species, i.e. the wavefunction yielding the lowest possible energy. Since the MC-QTAIM analysis of the derived local minima yields distinct atomic properties, it is hard to establish the true patterns of atomic property variations and they are not discussed in this contribution. New computational strategies are now under development in our lab to safely analyze this mass region.

The recently proposed extended theory of localization/delocalization of electrons and other quantum particles,[68] applicable within non-BO domain, was applied to the electrons of both series of species and final results are gathered in Tables 6 and 7. Since there is just a single PCP in the considered species, the localization/delocalization analysis is not applicable to the PCP.[68]

Inspection of both tables demonstrates that in line with previously observed trends the localization of electrons in the *X* basin as well as the delocalization of the electrons of the *X* basin into other basins all diminish. Comparison of the percent localization of the *X* basin with those of the nitrogen and carbon basins clearly reveals much smaller localization capacity of the *X* basin. However, the best probe of the discontinuous topological transformation is an abrupt increase in the localization index of the neighboring basin of the *X* basin when crossing the critical mass in both series of species. On the other hand, and in line with the nearsightedness principle, the electronic localization of basins that are not neighbors of the *X* basin as well as the delocalization of electrons between carbon and nitrogen basins are much less affected by the mass variation of the PCP and are almost constant in the whole mass spectrum.

## 4. Conclusions

In the present study through introducing the hybrid basis set and mobile functions in addition to the usual nuclear centered basis functions a large mass spectrum was scanned using the ab initio NEO-HF method. One may conceive the idea of mobile functions as in "fly" basis set design and this is particularly useful strategy when applying ab initio non-BO calculations to species with non-standard masses. However, even for real but less familiar quantum particles like the positive muon and associated muonic molecules, employing such strategy facilitates "from scratch" basis set design. Since this strategy may be used for both nuclear/PCP and electronic basis sets, the resulting basis functions are not biased, carrying the "fingerprint" of their "environment", which is a major advantage in contrast to the usual "pre-designed" basis sets with fixed variables. Thus, the large number of basis functions usually used to design flexible basis sets for quantum nuclei, could be bypassed and this is particularly desirable and a real save of computational cost when performing post-NEO-HF calculations. These issues will be addressed in detail in a future study.

The presented MC-QTAIM analysis reveals the detailed nature of the topological structural transformation upon the mass variation of the PCP, which in contrast to some similarities, is distinct from the usual geometry dependent topological structural transformations. Within the context of the orthodox QTAIM, the one-electron density parametrically depends on the position of clamped nuclei and the variations of MGs are accomplished by variations of molecular geometry.[15-17] These variations are confined to the rearrangement of a fixed number of AIM and atomic basins do not appear/disappear during geometrical variations except some very special cases,[88-90] or by applying extremely large hydrostatic pressure to moleucles.[91] However, this "AIM conservation" rule is restricted to the BO paradigm and present study demonstrates that within the context of the MC-QTAIM, beyond the BO paradigm, and upon mass variation of quantum particles, atomic basins may appear/disappear. Thus, the topological transformations considered in present study are a novel unprecedented type of structural transformations. It is timely to emphasize that idea of topological transformations have also been utilized recently to disclose the abrupt transition from $H^-$, atomic species, to $H_2^+$, molecular species, upon the mass variation of the constituent particle.[36,40] However, these topological transitions just disclose "topographical" changes of the used density and no underlying AIM structure was revealed in these studies.

In a very recent study the AIM structure of some very simple muonic species were considered within the context of the MC-QTAIM and it was proposed that the positive muon is capable of forming its own atomic basin.[74] In that study the positive muon competed with hydrogen isotopes, e.g. proton and deuteron, in shaping its own atomic basin. However, in this study it was demonstrated that in competition with the nitrogen atom in $CN$ moiety it is unable to shape an independent atomic basin and is absorbed into nitrogen basin. Accordingly, it seems legitimate to tentatively assign the following formulas, $\mu CN$ and $CN^-,\mu^+$, to the muonic species considered in present study in order to emphasize their structural resemblance to $HCN$ and $CN^-,e^+$, respectively; based on the ab initio calculations $\mu CN$ is the stable configuration. Before making a final decision on whether from the viewpoint of the AIM structure positive muon behaves like a lighter isotope of hydrogen or not, more MC-QTAIM studies on species containing the positive muon is needed. Accordingly, in a future contribution the MC-QTAIM analysis of a diverse set of muonic species will be considered comprehensively to shed some light on this interesting question.

It seems there is a consensus among theoretical chemists that the concept of chemical structure is applicable straightforwardly within the BO paradigm while it is not trivial to be applied in the non-BO domain.[18,42,92,93] Recent studies on analyzing non-BO wavefunctions shed light on how one may derive some ingredients of molecular structures in non-BO domain,[30-41] though serious technical obstacles yet remain to be tackled.[18,42] The present study, as well as the recent MC-QTAIM analysis of some polyatomic species,[70] however points that the clamped nucleus model is not required in order to derive the AIM structure which is one of the basic ingredients required to propose a chemical/molecular structure. More precisely, as far as there are some clamped nuclei in a molecule that the total translational and rotational motions are excluded from molecular non-BO wavefunction, assuming certain nuclei as quantum waves does not seem to be an obstacle to introduce molecular structures. Since chemists are usually interested in non-BO description of certain parts of a molecule,[47-63] e.g. isotope substitution in a specific site of a molecule, clamping some nuclei during ab initio non-BO calculations is not a real restriction and the NEO methodology may be applied successfully. The resulting non-BO wavefunctions are classified in a single category, and termed WF1 in a previous contribution,[65] while the MC-QTAIM analysis is capable of dealing with this class of wavefunctions yielding the underlying AIM structure. However, the subsequent question is: "What will happen if all nuclei are treated as quantum waves?". Nakai's proposed ab initio nuclear orbital plus molecular orbital method (NOMO) seems to be a powerful methodology in such cases which eliminates total translational and rotational motions systematically.[24] The resulting wavefunctions have been classified as WF2 and WF3 based on details of technicalities,[65] and after proper modifications of the present formalism of the MC-QTAIM, which will be discussed in a future contribution, the underlying AIM structure is also derivable from the NOMO wavefunctions. All these cast doubt that the BO paradigm and the clamped nucleus model are "the" border for applicability of the concept of molecular structure and more generally the Structural theory. However, at the same time, it must be stressed that for the most intricate non-BO wavefunctions that contain total rotational dynamics,[18] termed as WF4,[65] deriving the AIM structure is yet elusive,[40-42] and current MC-QTAIM methodology needs further theoretical developments to deal with such wavefunctions.

## Acknowledgments

The authors are grateful to Masumeh Gharabaghi, Cina Foroutan-Nejad, Rohoullah Firouzi and Shahin Sowlati for their detailed reading of a previous draft of this paper and helpful suggestions.

**Figure Legends**

**Fig. 1** The relief map of the Gamma density of selected species of the $XCN$ series of species when the masses of the PCPs are: $25m_e$ (a), $200m_e$ (b), $600m_e$ (c) and $1836m_e$ (d). The carbon nucleus is located in the center of the coordinate system while the nitrogen nucleus is located in the positive side of the z-axis.

**Fig. 2** The relief map of the Gamma density of selected species of the $CNX$ series of species when the masses of the PCPs are: $25m_e$ (a), $200m_e$ (b), $600m_e$ (c) and $1836m_e$ (d). The carbon nucleus is located in the negative side of the z-axis while the nitrogen nucleus is located in the center of the coordinate system.

**Fig. 3** The mass-scaled one-particle densities of the PCPs for two $XCN$ species including $m_X = 25m_e$ (the blue curve) and $m_X = m_e$ (the red curve). The positions of carbon and nitrogen nuclei are almost identical in both cases and are shown with green and orange circles, respectively.

Table 1- Some results of the ab initio calculations on the *XCN* and *CNX* series of species including *C-N* (*N-C*) inter-nuclear distances, *C-X* and *N-X* mean inter-nuclear distances, total energies as well as virial ratios. All results are given in atomic units.

| *XCN* *X-mass* | N-C | C-X* | Energy | virial ratio | *CNX* *X-mass* | C-N | N-X* | Energy | virial ratio |
|---|---|---|---|---|---|---|---|---|---|
| *1* | 2.169 | -- | -92.4612 | 2.0003 | *25* | 2.175 | -- | -92.6455 | 2.0013 |
| *25* | 2.136 | -- | -92.6609 | 2.0007 | *50* | 2.173 | -- | -92.6940 | 2.0013 |
| *50* | 2.134 | -- | -92.7105 | 2.0007 | *100* | 2.173 | -- | -92.7350 | 2.0010 |
| *60* | 2.134 | -- | -92.7222 | 2.0007 | *200* | 2.172 | -- | -92.7681 | 2.0011 |
| *65* | 2.133 | -- | -92.7272 | 2.0007 | *300* | 2.171 | -- | -92.7843 | 2.0011 |
| *70* | 2.133 | -- | -92.7317 | 2.0007 | *350* | 2.171 | -- | -92.7898 | 2.0011 |
| *75* | 2.133 | -- | -92.7358 | 2.0007 | *385* | 2.171 | -- | -92.7926 | 2.0012 |
| *80* | 2.133 | -- | -92.7396 | 2.0007 | *400* | 2.171 | -- | -92.7939 | 2.0012 |
| *85* | 2.133 | 2.222 | -92.7431 | 2.0007 | *405* | 2.171 | -- | -92.7943 | 2.0012 |
| *90* | 2.133 | 2.216 | -92.7463 | 2.0007 | *410* | 2.171 | 1.953 | -92.7947 | 2.0012 |
| *95* | 2.133 | 2.210 | -92.7493 | 2.0007 | *415* | 2.171 | 1.953 | -92.7951 | 2.0012 |
| *100* | 2.133 | 2.205 | -92.7521 | 2.0007 | *425* | 2.171 | 1.952 | -92.7959 | 2.0012 |
| *200* | 2.132 | 2.145 | -92.7860 | 2.0007 | *500* | 2.171 | 1.944 | -92.8011 | 2.0012 |
| *600* | 2.131 | 2.086 | -92.8254 | 2.0007 | *600* | 2.171 | 1.937 | -92.8065 | 2.0012 |
| *1000* | 2.130 | 2.067 | -92.8387 | 2.0007 | *1000* | 2.171 | 1.919 | -92.8196 | 2.0012 |
| *1400* | 2.130 | 2.058 | -92.8462 | 2.0007 | *1400* | 2.170 | 1.909 | -92.8270 | 2.0012 |
| *1836* | 2.130 | 2.051 | -92.8514 | 2.0007 | *1836* | 2.170 | 1.903 | -92.8322 | 2.0012 |

* The mean inter-nuclear distance is the distance between the clamped nucleus and the center of the s-type nuclear Gaussian function describing the quantum nucleus. For $m_X \leq 80 m_e$ in the *XCN* series and $m_X \leq 405 m_e$ in the *CNX* series the atomic basin corresponding to *X* particle disappears thus the mean inter-nuclear distances are not reported.

Table 2- Some results of the topological analysis on the *XCN* and *CNX* series of species including the Gamma density, the combined Lagrangian kinetic energy density (denoted as G), the Laplacian of the Gamma density, computed (3, -3) CP located in *X* basin and at the (3, -1) linking the (3, -3) CP on carbon nucleus and the (3, -3) CP within the *X* basin for the *XCN* series and at the (3, -1) CP linking the (3, -3) CP on nitrogen nucleus and the (3, -3) CP within the *X* basin for the *CNX* series. The length of the path connecting the mentioned CPs (denoted as L(X-(3, -1))), and the index of the topological floppiness (denoted as TF) are also presented. All results are given in atomic units.

| XCN | Gamma | | G | | Laplacian of Gamma | | | |
|---|---|---|---|---|---|---|---|---|
| X-mass | (3, -1) | X-(3, -3) | (3, -1) | X-(3, -3) | (3, -1) | X-(3, -3) | L (X-(3, -1)) | TF |
| *85* | 0.197 | 0.197 | 0.047 | 0.045 | -1.363 | -1.483 | 0.028 | 1.0000 |
| *100* | 0.204 | 0.205 | 0.053 | 0.042 | -1.145 | -2.018 | 0.176 | 0.9944 |
| *200* | 0.228 | 0.239 | 0.045 | 0.045 | -1.074 | -3.746 | 0.339 | 0.9538 |
| *600* | 0.255 | 0.286 | 0.013 | 0.061 | -1.112 | -7.220 | 0.476 | 0.8914 |
| *1000* | 0.263 | 0.304 | 0.013 | 0.071 | -1.067 | -9.172 | 0.516 | 0.8656 |
| *1400* | 0.267 | 0.314 | 0.013 | 0.079 | -1.047 | -10.564 | 0.533 | 0.8505 |
| *1836* | 0.270 | 0.322 | 0.014 | 0.087 | -1.039 | -11.761 | 0.544 | 0.8397 |
| **NCX** | | | | | | | | |
| *410* | 0.269 | 0.269 | 0.128 | 0.127 | -3.280 | -3.416 | 0.008 | 1.0000 |
| *425* | 0.271 | 0.271 | 0.132 | 0.124 | -3.024 | -3.787 | 0.042 | 0.9998 |
| *500* | 0.276 | 0.277 | 0.138 | 0.122 | -2.740 | -4.617 | 0.096 | 0.9976 |
| *600* | 0.282 | 0.284 | 0.138 | 0.124 | -2.631 | -5.382 | 0.128 | 0.9942 |
| *1000* | 0.297 | 0.302 | 0.117 | 0.138 | -2.642 | -7.371 | 0.178 | 0.9842 |
| *1400* | 0.306 | 0.312 | 0.084 | 0.150 | -2.729 | -8.726 | 0.207 | 0.9781 |
| *1836* | 0.312 | 0.320 | 0.052 | 0.162 | -2.745 | -9.850 | 0.237 | 0.9733 |

Table 3- The results of the basin integration of the *X* basin in the *XCN* and *CNX* series of species including basin energy, electronic population (donated as e-pop.), the PCP population (denoted as PCP-pop.) and the volume of the atomic basin. All results are given in atomic units.

| *XCN* | | | | |
|---|---|---|---|---|
| *X-mass* | **Energy** | **e-pop.** | **PCP-pop.** | **Volume** |
| *85* | -0.2212 | 0.325 | 0.856 | 24.0 |
| *100* | -0.2514 | 0.375 | 0.915 | 26.4 |
| *200* | -0.3083 | 0.462 | 0.984 | 29.3 |
| *600* | -0.3824 | 0.582 | 1.000 | 32.8 |
| *1000* | -0.4074 | 0.621 | 1.000 | 33.9 |
| *1400* | -0.4199 | 0.640 | 1.000 | 34.4 |
| *1836* | -0.4284 | 0.652 | 1.000 | 34.8 |
| *NCX* | | | | |
| *410* | -0.1968 | 0.220 | 0.863 | 14.0 |
| *425* | -0.2053 | 0.231 | 0.880 | 14.3 |
| *500* | -0.2200 | 0.249 | 0.919 | 15.1 |
| *600* | -0.2305 | 0.261 | 0.944 | 15.6 |
| *1000* | -0.2518 | 0.286 | 0.981 | 16.4 |
| *1400* | -0.2669 | 0.305 | 0.993 | 17.1 |
| *1836* | -0.2823 | 0.326 | 0.998 | 17.8 |

Table 4- The results of the basin integration of the nitrogen and carbon basins in the *XCN* series of species including basin energies, electronic populations (donated as e-pop.), the PCP populations (denoted as PCP-pop.) and the volumes of the atomic basins. The line between $m_X = 80m_e$ and $m_X = 85m_e$ is the border between triatomic and diatomic structures. All results are given in atomic units.

| X-mass | N-basin | | | | C-basin | | | |
|---|---|---|---|---|---|---|---|---|
| | Energy | e-pop. | PCP-pop. | Volume | Energy | e-pop. | PCP-pop. | Volume |
| *1* | -55.5023 | 8.874 | 0.618 | 276.1 | -36.9590 | 5.126 | 0.382 | 203.7 |
| *25* | -55.3891 | 8.522 | 0.000 | 168.9 | -37.2719 | 5.478 | 1.000 | 140.9 |
| *50* | -55.3838 | 8.510 | 0.000 | 168.5 | -37.3268 | 5.490 | 1.000 | 134.9 |
| *60* | -55.3828 | 8.508 | 0.000 | 168.4 | -37.3393 | 5.493 | 1.000 | 133.7 |
| *65* | -55.3817 | 8.506 | 0.000 | 168.3 | -37.3455 | 5.494 | 1.000 | 133.3 |
| *70* | -55.3809 | 8.506 | 0.000 | 168.4 | -37.3507 | 5.495 | 1.000 | 132.4 |
| *75* | -55.3804 | 8.505 | 0.000 | 167.6 | -37.3554 | 5.495 | 1.000 | 131.9 |
| *80* | -55.3799 | 8.505 | 0.000 | 167.5 | -37.3597 | 5.495 | 1.000 | 131.5 |
| *85* | -55.3795 | 8.504 | 0.000 | 167.5 | -37.1424 | 5.171 | 0.145 | 107.0 |
| *100* | -55.3784 | 8.503 | 0.000 | 167.5 | -37.1223 | 5.122 | 0.085 | 104.1 |
| *200* | -55.3754 | 8.499 | 0.000 | 167.2 | -37.1022 | 5.039 | 0.016 | 98.6 |
| *600* | -55.3725 | 8.495 | 0.000 | 167.1 | -37.0704 | 4.923 | 0.000 | 92.3 |
| *1000* | -55.3716 | 8.494 | 0.000 | 167.0 | -37.0598 | 4.884 | 0.000 | 89.9 |
| *1400* | -55.3712 | 8.494 | 0.000 | 167.0 | -37.0551 | 4.866 | 0.000 | 88.6 |
| *1836* | -55.3708 | 8.494 | 0.000 | 167.0 | -37.0522 | 4.854 | 0.000 | 87.9 |

Table 5- The results of the basin integration of the nitrogen and carbon basins in the *CNX* series of species including basin energies, electronic populations (donated as e-pop.), the PCP populations (denoted as PCP-pop.) and the volumes of the atomic basins. The line between $m_X = 405m_e$ and $m_X = 410m_e$ is the border between triatomic and diatomic structures. All results are given in atomic units.

| | C-basin | | | | N-basin | | | |
|---|---|---|---|---|---|---|---|---|
| *X-mass* | **Energy** | **e-pop.** | **PCP-pop.** | **Volume** | **Energy** | **e-pop.** | **PCP-pop.** | **Volume** |
| *25* | -36.8621 | 4.822 | 0.000 | 121.1 | -55.7833 | 9.178 | 1.000 | 185.8 |
| *50* | -36.8588 | 4.812 | 0.000 | 120.2 | -55.8351 | 9.188 | 1.000 | 182.1 |
| *100* | -36.8575 | 4.807 | 0.000 | 119.7 | -55.8775 | 9.193 | 1.000 | 179.5 |
| *200* | -36.8466 | 4.805 | 0.000 | 119.7 | -55.9215 | 9.195 | 1.000 | 177.9 |
| *300* | -36.8455 | 4.803 | 0.000 | 119.5 | -55.9387 | 9.197 | 1.000 | 177.4 |
| *350* | -36.8449 | 4.802 | 0.000 | 119.5 | -55.9449 | 9.198 | 1.000 | 177.3 |
| *385* | -36.8535 | 4.799 | 0.000 | 118.3 | -55.9391 | 9.201 | 1.000 | 177.2 |
| *400* | -36.8535 | 4.799 | 0.000 | 118.2 | -55.9404 | 9.201 | 1.000 | 177.2 |
| *405* | -36.8535 | 4.799 | 0.000 | 118.2 | -55.9408 | 9.201 | 1.000 | 177.2 |
| *410* | -36.8535 | 4.799 | 0.000 | 118.2 | -55.7444 | 8.981 | 0.137 | 163.2 |
| *425* | -36.8534 | 4.799 | 0.000 | 118.1 | -55.7371 | 8.970 | 0.121 | 162.7 |
| *500* | -36.8534 | 4.798 | 0.000 | 117.9 | -55.7277 | 8.953 | 0.081 | 161.8 |
| *600* | -36.8532 | 4.798 | 0.000 | 117.8 | -55.7227 | 8.941 | 0.056 | 161.2 |
| *1000* | -36.8500 | 4.797 | 0.000 | 118.4 | -55.7178 | 8.917 | 0.019 | 158.9 |
| *1400* | -36.8499 | 4.797 | 0.000 | 118.4 | -55.7101 | 8.898 | 0.007 | 157.7 |
| *1836* | -36.8498 | 4.796 | 0.000 | 118.4 | -55.7000 | 8.878 | 0.002 | 156.6 |

Table 6- The electronic localization and delocalization indices as well as the percent localization of the $XCN$ series of species. The line between $m_X = 80 m_e$ and $m_X = 85 m_e$ is the border between triatomic and diatomic structures.

| X-mass | Loc. | | | % Loc. | | | Deloc. | | |
|---|---|---|---|---|---|---|---|---|---|
| | N | C | X | N | C | X | (N,C) | (X,C) | (N,X) |
| *1* | 7.81 | 4.06 | -- | 88.0 | 79.2 | -- | 2.13 | -- | -- |
| *25* | 7.37 | 4.32 | -- | 86.4 | 78.9 | -- | 2.31 | -- | -- |
| *50* | 7.35 | 4.33 | -- | 86.4 | 78.9 | -- | 2.31 | -- | -- |
| *60* | 7.35 | 4.34 | -- | 86.4 | 78.9 | -- | 2.31 | -- | -- |
| *65* | 7.35 | 4.34 | -- | 86.4 | 78.9 | -- | 2.31 | -- | -- |
| *70* | 7.35 | 4.32 | -- | 86.4 | 78.6 | -- | 2.31 | -- | -- |
| *75* | 7.35 | 4.32 | -- | 86.4 | 75.7 | -- | 2.31 | -- | -- |
| *80* | 7.35 | 4.32 | -- | 86.4 | 75.7 | -- | 2.30 | -- | -- |
| 85 | 7.35 | 3.76 | 0.05 | 86.4 | 72.9 | 16.1 | 2.28 | 0.52 | 0.03 |
| 100 | 7.35 | 3.70 | 0.07 | 86.4 | 72.1 | 18.2 | 2.28 | 0.58 | 0.04 |
| *200* | 7.34 | 3.57 | 0.10 | 86.4 | 70.8 | 22.3 | 2.27 | 0.67 | 0.04 |
| *600* | 7.34 | 3.40 | 0.16 | 86.4 | 69.1 | 27.9 | 2.26 | 0.78 | 0.06 |
| *1000* | 7.34 | 3.35 | 0.18 | 86.4 | 68.6 | 29.7 | 2.25 | 0.81 | 0.06 |
| *1400* | 7.34 | 3.33 | 0.20 | 86.4 | 68.4 | 30.5 | 2.25 | 0.83 | 0.06 |
| *1836* | 7.34 | 3.31 | 0.20 | 86.4 | 68.2 | 31.0 | 2.25 | 0.84 | 0.06 |

Table 7- The electronic localization and delocalization indices as well as the percent localization of the $CNX$ series of species. The line between $m_X = 405m_e$ and $m_X = 410m_e$ is the border between triatomic and diatomic structures.

|  | Loc. | | | % Loc. | | | Deloc. | | |
|---|---|---|---|---|---|---|---|---|---|
| X-mass | C | N | X | C | N | X | (C,N) | (C,X) | (N,X) |
| 25 | 3.93 | 8.29 | -- | 81.5 | 90.3 | -- | 1.78 | -- | -- |
| 50 | 3.93 | 8.31 | -- | 81.7 | 90.4 | -- | 1.77 | -- | -- |
| 100 | 3.93 | 8.32 | -- | 81.8 | 90.5 | -- | 1.75 | -- | -- |
| 200 | 3.93 | 8.32 | -- | 81.8 | 90.5 | -- | 1.75 | -- | -- |
| 300 | 3.93 | 8.32 | -- | 81.8 | 90.5 | -- | 1.75 | -- | -- |
| 350 | 3.93 | 8.32 | -- | 81.8 | 90.5 | -- | 1.75 | -- | -- |
| 385 | 3.93 | 8.32 | -- | 81.8 | 90.5 | -- | 1.74 | -- | -- |
| 405 | 3.93 | 8.34 | -- | 81.8 | 90.6 | -- | 1.74 | -- | -- |
| 410 | 3.93 | 7.92 | 0.02 | 81.8 | 88.2 | 11.1 | 1.73 | 0.01 | 0.39 |
| 425 | 3.93 | 7.90 | 0.03 | 81.8 | 88.1 | 11.2 | 1.73 | 0.01 | 0.40 |
| 500 | 3.93 | 7.87 | 0.03 | 81.8 | 88.0 | 11.9 | 1.73 | 0.01 | 0.43 |
| 600 | 3.93 | 7.85 | 0.03 | 81.8 | 87.8 | 12.5 | 1.73 | 0.01 | 0.44 |
| 1000 | 3.93 | 7.81 | 0.04 | 81.9 | 87.6 | 13.7 | 1.73 | 0.01 | 0.48 |
| 1400 | 3.93 | 7.78 | 0.04 | 81.9 | 87.5 | 14.6 | 1.72 | 0.01 | 0.51 |
| 1836 | 3.93 | 7.75 | 0.05 | 81.9 | 87.3 | 15.5 | 1.72 | 0.02 | 0.54 |

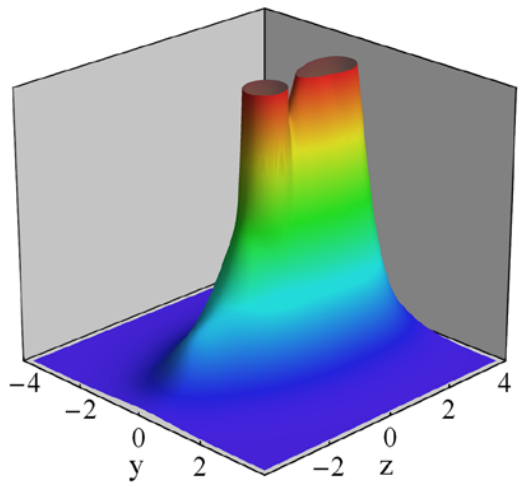
(a)

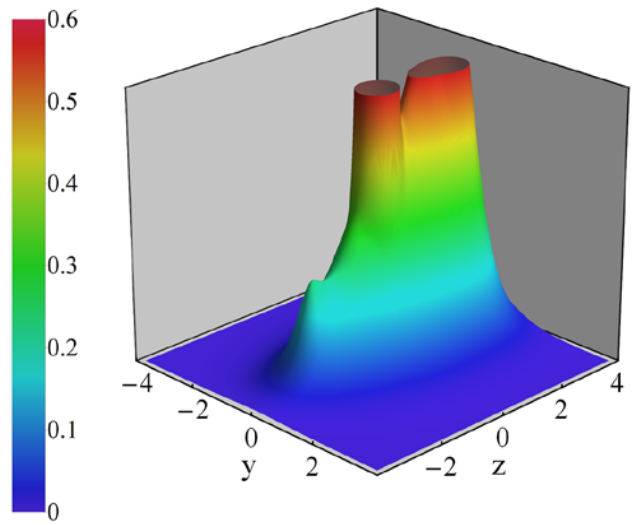
(b)

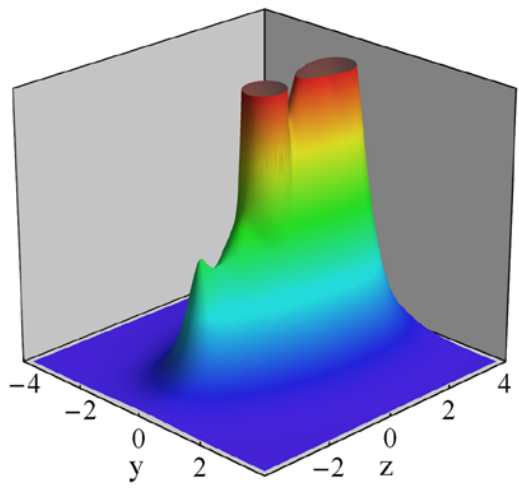
(c)

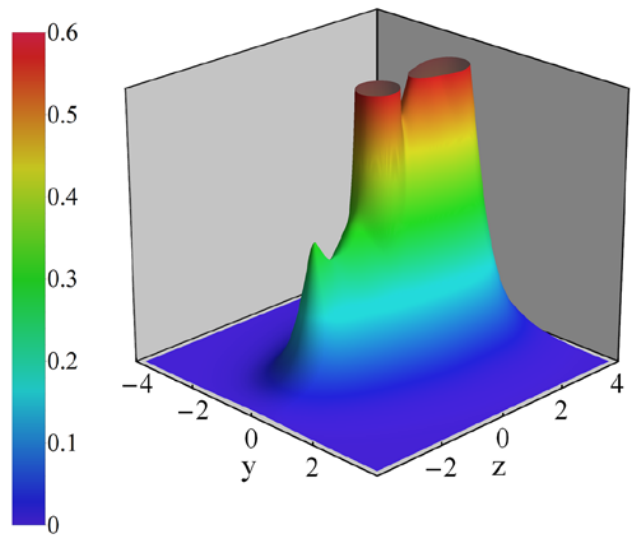
(d)

**Figure-1**

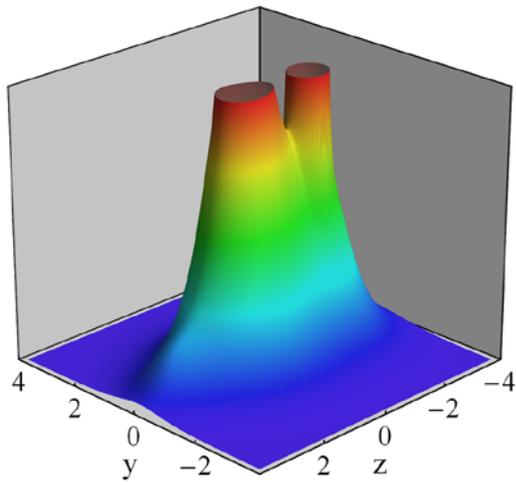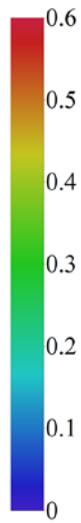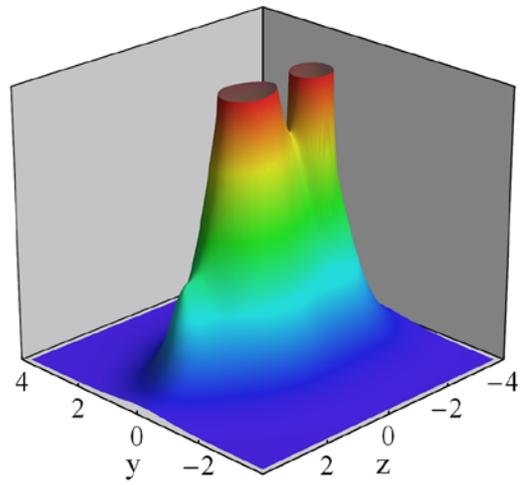

(a) (b)

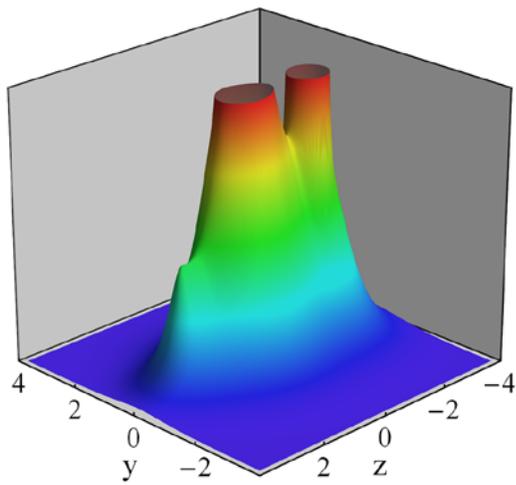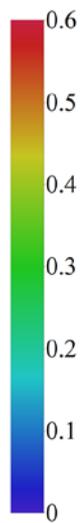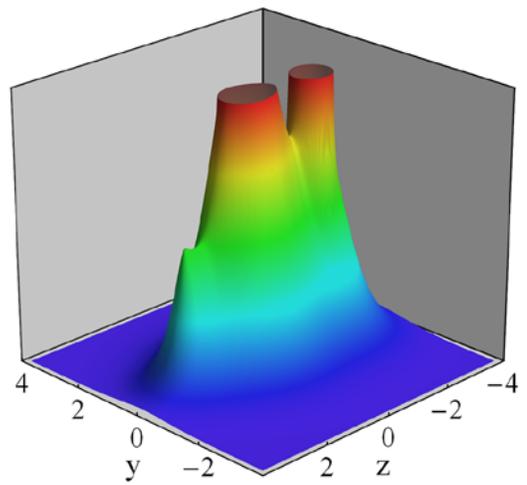

(c) (d)

**Figure-2**

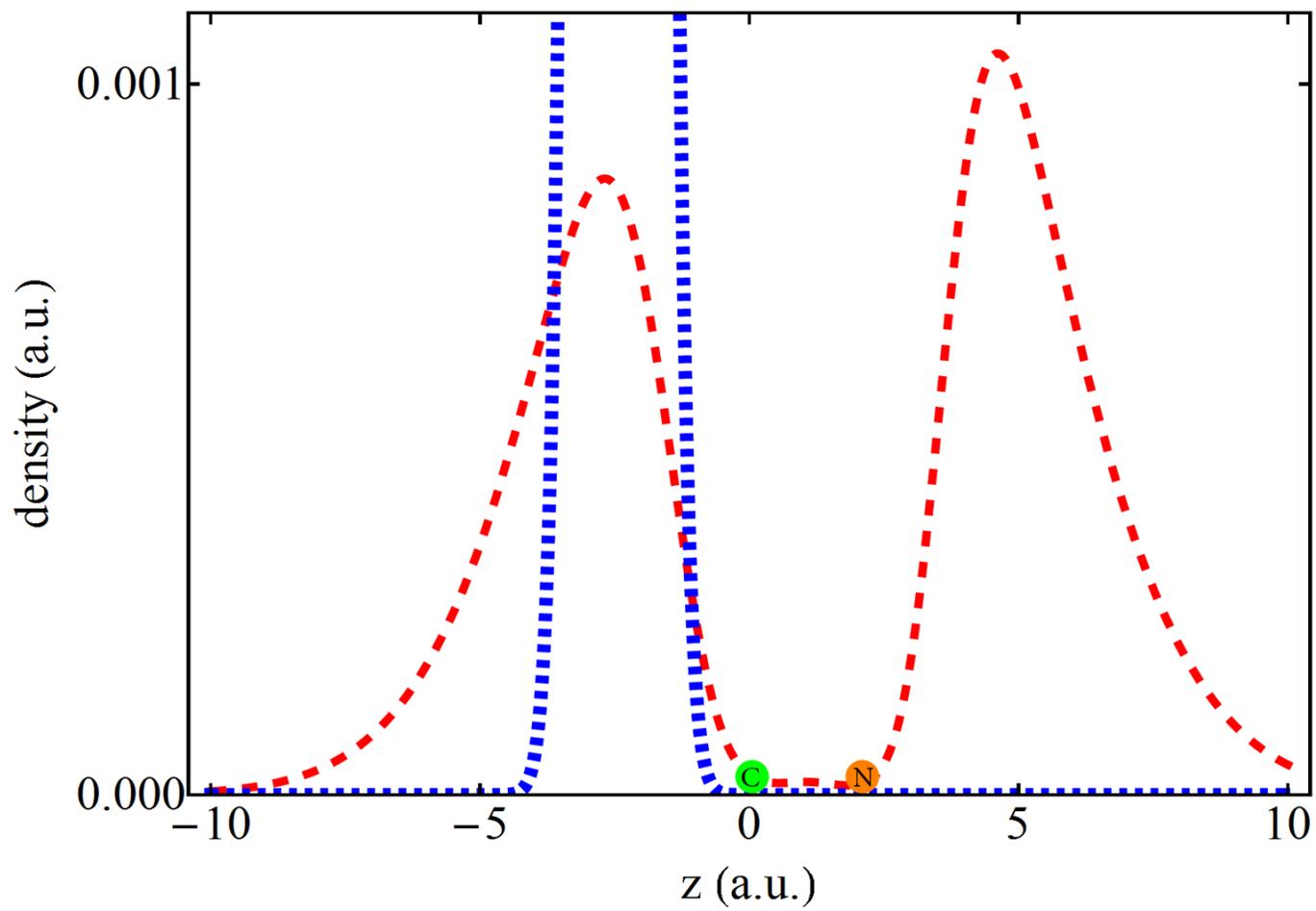

**Figue-3**

# Supporting Information

## Hidden aspects of the Structural Theory of chemistry: The MC-QTAIM analysis reveals "alchemical" transformation from a triatomic to a diatomic structure

Mohammad Goli and Shant Shahbazian[*]

*Faculty of Chemistry, Shahid Beheshti University, G. C. , Evin, Tehran, Iran, 19839, P.O. Box 19395-4716.*

Tel/Fax: 98-21-22431661

E-mail:
(Shant Shahbazian) chemist_shant@yahoo.com

[*] Corresponding author

# Table of contents

Table S1- Some results of the topological analysis on the *XCN* series of species including the Gamma density, the combined Lagrangian kinetic energy density (denoted as G), the Laplacian of the Gamma density (denoted as Lap. Γ), computed at the (3, -3) CP on carbon and nitrogen clamped nuclei and the (3, -1) CP connecting these two CPs. The line between $m_X = 80m_e$ and $m_X = 85m_e$ is the border between triatomic and diatomic structures. All results are given in atomic units.

| *X-mass* | *Gamma* | | | G | | | Lap. Γ | | |
|---|---|---|---|---|---|---|---|---|---|
| | **N-(3, -3)** | **LCP** | **C-(3,-3)** | **N-(3, -3)** | **LCP** | **C-(3,-3)** | **N-(3, -3)** | **LCP** | **C-(3,-3)** |
| *1* | 196.5 | 0.520 | 121.4 | 17.1 | 1.046 | 5.0 | -1476301 | -0.090 | -670837 |
| *25* | 196.9 | 0.522 | 120.9 | 17.0 | 1.271 | 5.4 | -1480090 | 0.998 | -667666 |
| *50* | 196.9 | 0.523 | 120.9 | 17.0 | 1.276 | 5.5 | -1480078 | 1.011 | -667525 |
| *60* | 196.9 | 0.523 | 120.9 | 17.0 | 1.276 | 5.5 | -1480074 | 1.011 | -667487 |
| *65* | 196.9 | 0.523 | 120.9 | 17.0 | 1.276 | 5.5 | -1480030 | 1.007 | -667423 |
| *70* | 196.9 | 0.523 | 120.9 | 17.0 | 1.276 | 5.5 | -1480016 | 1.008 | -667417 |
| *75* | 196.9 | 0.523 | 120.9 | 17.0 | 1.276 | 5.5 | -1480008 | 1.008 | -667414 |
| *80* | 196.9 | 0.523 | 120.9 | 17.0 | 1.276 | 5.5 | -1480004 | 1.008 | -667411 |
| *85* | 196.9 | 0.523 | 120.9 | 17.0 | 1.276 | 5.5 | -1479999 | 1.009 | -667409 |
| *100* | 196.9 | 0.523 | 120.9 | 17.0 | 1.276 | 5.5 | -1479990 | 1.010 | -667406 |
| *200* | 196.9 | 0.524 | 120.9 | 17.0 | 1.278 | 5.5 | -1479966 | 1.015 | -667399 |
| *600* | 196.9 | 0.524 | 120.9 | 17.0 | 1.279 | 5.6 | -1479945 | 1.018 | -667412 |
| *1000* | 196.9 | 0.524 | 120.9 | 17.0 | 1.280 | 5.6 | -1479935 | 1.019 | -667424 |
| *1400* | 196.9 | 0.524 | 120.9 | 17.0 | 1.280 | 5.6 | -1479931 | 1.019 | -667434 |
| *1836* | 196.9 | 0.524 | 120.9 | 17.0 | 1.280 | 5.6 | -1479929 | 1.019 | -667441 |

Table S2- Some results of the topological analysis on the *CNX* series of species including the Gamma density, the combined Lagrangian kinetic energy density (denoted as G), the Laplacian of the Gamma density (denoted as Lap. Γ), computed at the (3, -3) CP on carbon and nitrogen clamped nuclei and the (3, -1) CP connecting these two CPs. The line between $m_X = 405 m_e$ and $m_X = 410 m_e$ is the border between triatomic and diatomic structures. All results are given in atomic units.

| | *Gamma* | | | G | | | Lap. Γ | | |
|---|---|---|---|---|---|---|---|---|---|
| *X-mass* | C-(3,-3) | LCP | N-(3, -3) | C-(3,-3) | LCP | N-(3, -3) | C-(3,-3) | LCP | N-(3, -3) |
| *25* | 121.6 | 0.488 | 195.7 | 4.7 | 1.015 | 18.1 | -671651 | 0.210 | -1469770 |
| *50* | 121.6 | 0.486 | 195.6 | 4.6 | 1.016 | 18.2 | -671610 | 0.245 | -1469217 |
| *100* | 121.7 | 0.481 | 195.7 | 4.6 | 1.019 | 18.3 | -672369 | 0.330 | -1470200 |
| *200* | 121.6 | 0.482 | 195.7 | 4.6 | 1.016 | 18.4 | -671676 | 0.301 | -1470216 |
| *300* | 121.6 | 0.481 | 195.7 | 4.6 | 1.016 | 18.4 | -671648 | 0.309 | -1470135 |
| *350* | 121.6 | 0.481 | 195.7 | 4.6 | 1.016 | 18.4 | -671629 | 0.312 | -1470117 |
| *385* | 121.6 | 0.479 | 195.6 | 4.5 | 1.045 | 18.4 | -671897 | 0.487 | -1468964 |
| *400* | 121.6 | 0.479 | 195.6 | 4.5 | 1.045 | 18.4 | -671893 | 0.488 | -1468955 |
| *405* | 121.6 | 0.479 | 195.6 | 4.5 | 1.045 | 18.4 | -671893 | 0.488 | -1468952 |
| *410* | 121.6 | 0.479 | 195.6 | 4.5 | 1.045 | 18.4 | -671894 | 0.489 | -1468950 |
| *425* | 121.6 | 0.479 | 195.6 | 4.5 | 1.045 | 18.5 | -671893 | 0.489 | -1468940 |
| *500* | 121.6 | 0.478 | 195.6 | 4.5 | 1.045 | 18.5 | -671896 | 0.492 | -1468903 |
| *600* | 121.6 | 0.478 | 195.6 | 4.5 | 1.045 | 18.5 | -671893 | 0.496 | -1468862 |
| *1000* | 121.6 | 0.478 | 195.6 | 4.5 | 1.043 | 18.5 | -671906 | 0.502 | -1468815 |
| *1400* | 121.6 | 0.477 | 195.6 | 4.5 | 1.044 | 18.5 | -671911 | 0.506 | -1468767 |
| *1836* | 121.6 | 0.477 | 195.6 | 4.5 | 1.044 | 18.5 | -671916 | 0.509 | -1468737 |

Table S3- The separate electronic and PCP contributions to the Gamma density, the combined Lagrangian kinetic energy density (denoted as G), the Laplacian of the Gamma density (denoted as Lap. Γ) all computed at the (3, -3) CP located in the *X* basin and at the (3, -1) linking the (3, -3) CP on carbon nucleus and the (3, -3) CP within the *X* basin the for the *XCN* series of species. All results are given in atomic units.

|  | Gamma[*] | | G | | Laplacian of Gamma[*] | |
|---|---|---|---|---|---|---|
| **X-mass** | (3, -1) | X-(3, -3) | (3, -1) | X-(3, -3) | (3, -1) | X-(3, -3) |
| **electronic** | | | | | | |
| *85* | 0.181 | 0.179 | 0.007 | 0.007 | -1.061 | -1.087 |
| *100* | 0.194 | 0.182 | 0.007 | 0.007 | -1.071 | -1.288 |
| *200* | 0.226 | 0.211 | 0.007 | 0.008 | -1.207 | -2.099 |
| *600* | 0.255 | 0.257 | 0.011 | 0.008 | -1.125 | -3.682 |
| *1000* | 0.263 | 0.276 | 0.012 | 0.009 | -1.067 | -4.498 |
| *1400* | 0.267 | 0.288 | 0.013 | 0.009 | -1.047 | -5.046 |
| *1836* | 0.270 | 0.296 | 0.014 | 0.009 | -1.037 | -5.481 |
| **PCP** | | | | | | |
| *85* | 0.016 | 0.018 | 0.040 | 0.038 | -0.302 | -0.396 |
| *100* | 0.010 | 0.023 | 0.046 | 0.035 | -0.074 | -0.730 |
| *200* | 0.002 | 0.028 | 0.038 | 0.038 | 0.133 | -1.647 |
| *600* | 0.000 | 0.028 | 0.002 | 0.052 | 0.013 | -3.538 |
| *1000* | 0.000 | 0.027 | 0.000 | 0.062 | 0.001 | -4.674 |
| *1400* | 0.000 | 0.026 | 0.000 | 0.070 | 0.000 | -5.519 |
| *1836* | 0.000 | 0.026 | 0.000 | 0.078 | 0.000 | -6.249 |

[*] The PCP's contribution to the Gamma density and its Laplacian is the mass scaled one-particle density and its Laplacian (see section 2 of paper for details).

Table S4- The separate electronic and PCP contributions to the Gamma density, the combined Lagrangian kinetic energy density (denoted as G), the Laplacian of the Gamma density (denoted as Lap. Γ) all computed at the (3, -3) CP located in the *X* basin and at the (3, -1) linking the (3, -3) CP on nitrogen nucleus and the (3, -3) CP within the *X* basin for the *CNX* series of species. All results are given in atomic units.

| X-mass | Gamma* | | G | | Laplacian of Gamma* | |
|---|---|---|---|---|---|---|
| | (3, -1) | X-(3, -3) | (3, -1) | X-(3, -3) | (3, -1) | X-(3, -3) |
| **electronic** | | | | | | |
| *410* | 0.255 | 0.254 | 0.018 | 0.018 | -2.566 | -2.594 |
| *425* | 0.258 | 0.253 | 0.018 | 0.018 | -2.544 | -2.697 |
| *500* | 0.267 | 0.257 | 0.018 | 0.018 | -2.605 | -3.005 |
| *600* | 0.276 | 0.263 | 0.018 | 0.018 | -2.703 | -3.328 |
| *1000* | 0.294 | 0.280 | 0.018 | 0.019 | -2.961 | -4.208 |
| *1400* | 0.304 | 0.291 | 0.019 | 0.019 | -3.021 | -4.801 |
| *1836* | 0.311 | 0.299 | 0.020 | 0.019 | -2.913 | -5.279 |
| **PCP** | | | | | | |
| *410* | 0.015 | 0.016 | 0.110 | 0.108 | -0.713 | -0.823 |
| *425* | 0.013 | 0.017 | 0.114 | 0.105 | -0.481 | -1.090 |
| *500* | 0.009 | 0.020 | 0.120 | 0.104 | -0.135 | -1.611 |
| *600* | 0.006 | 0.021 | 0.120 | 0.106 | 0.071 | -2.054 |
| *1000* | 0.003 | 0.022 | 0.099 | 0.119 | 0.319 | -3.163 |
| *1400* | 0.001 | 0.022 | 0.066 | 0.131 | 0.292 | -3.925 |
| *1836* | 0.000 | 0.021 | 0.032 | 0.143 | 0.168 | -4.570 |

* The PCP's contribution to the Gamma density and its Laplacian is the mass scaled one-particle density and its Laplacian (see section 2 of paper for details).